\documentclass{elsart}
\usepackage{amssymb}
\usepackage{epsfig,epsf,graphicx}
\usepackage{amsmath}
\def\be{\begin{eqnarray} &&}

\def\ee{\end{eqnarray}}

\def\bew{\begin{widetext}}
\def\ew{\end{widetext}}

\begin{document}
\begin{frontmatter}

\title{Linking Dynamical Gluon Mass to Chiral Symmetry Breaking via a QCD Low Energy  Effective Field Theory}

\author[ITA,COI]{O. Oliveira}, \author[ITA]{W. de Paula} and \author[ITA]{T. Frederico}
\address[ITA]{Departamento de F\'\i sica, Instituto Tecnol\'ogico de Aeron\'autica, 12228-900 S\~ao Jos\'e dos Campos, SP, Brazil.}

\address[COI]{Departamento de F\'{\i}sica, Universidade de Coimbra, 3004-516 Coimbra, Portugal}
\date{\today}
\maketitle

\begin{abstract}
A low energy effective field theory model for QCD with a scalar
color octet field is discussed. The model relates the gluon mass, the constituent quark masses
and the quark condensate. The gluon mass comes about $\sqrt{N_c}\, \Lambda_{QCD}$
with the quark condensate being proportional to the gluon mass squared. The model suggests
that the restoration of chiral symmetry and the deconfinement transition occur at the same temperature
and that, near the transition, the critical exponent for the condensate is twice the gluon mass one.
The model also favors the decoupling like solution for the gluon propagator.
\end{abstract}
\end{frontmatter}

%====================================================================
%====================================================================

The use of effective field theories is a common practice in Physics (see e.g. \cite{EffectiveTheory}).
There are many examples in the literature of such type of approach for low energy hadronic
physics from chiral perturbation theory (see e.g. \cite{ChiralPertTheory}), to
quark models with contact interactions and quark models including
pions and meson fields (see e.g.\cite{QuarkModels}).
In this work, we  discuss an effective model, invariant under local
$SU(3)$ color, for the non-per\-tur\-ba\-ti\-ve regime of QCD which uses as degrees of freedom
the gluon, the quark  and a color octet scalar field $\phi^a$.
The scalar field can be viewed as multi-gluon and multi-quark excitation of the vacuum, i.e.
it should be interpreted as a
collective excitation of the QCD vacuum. Given that the scalar field is not a color singlet, $\phi^a$ does not
contribute directly to the S-matrix. The scalar field resumes the
non-perturbative properties of QCD and there is no need to specify in detail its dynamics.
The $\phi^a$ contribution to the quark and gluon dynamics comes via
the color singlet condensate $\langle \phi^a \phi^a \rangle$ which,
according to our estimates, turns out to be of the order of
$\Lambda_{QCD}$.
The scalar field generate a gluon mass $m_g$, shifts the current
quark mass  and contributes to the quark
condensate $\langle \overline q q\rangle$. It turns out that the
shift in the quark mass and the quark condensate are all proportional to
the effective gluon mass. In this sense, the model links the
presence of an effective gluon mass with chiral symmetry breaking ($\chi S B$).

The studies of the phase diagram of QCD \cite{Costa2009}
suggest that the restoration of chiral symmetry and the deconfinement phase transition occur
at similar values for the temperature and chemical potential, at least for small
enough values of the chemical potential $\mu \lesssim 300$ MeV. In
the model under discussion either the gluon gets a dynamically
generated mass, quarks are confined and chiral symmetry is broken or
the gluon mass vanish, quarks become deconfined and chiral symmetry
is restored. The model predicts that the
restoration of chiral symmetry and deconfinement transition occur simultaneously at low baryon density,
in close agreement with the study of the QCD phase-diagram.

Some of the results  obtained here can be tested comparing the
predictions with non-perturbative solutions of QCD, namely with the
results from Schwinger-Dyson equations (SDE) and lattice
simulations.
For pure gauge, the recent solutions of the SDE can be classified into two classes. The scaling
solution \cite{SDEscaling}
which predicts a
vanishing gluon propagator at zero momenta and has no connection
with a dynamical generated gluon mass. The decoupling solution
\cite{Cornwall1982,Cornwall2009,Aguilar2011}
which describes the gluon as a massive particle in the low energy
regime, with the non-vanishing mass being related with gluon
confinement. Other authors have also contributed to the
understanding of the solutions of the SDE, see
\cite{Maris2003}.
The two solutions agree in the ultraviolet region but differ
significantly in the infrared region, with the decoupling solution
being closer to the results of lattice QCD simulations.

Lattice QCD  simulations \cite{Oliveira2011} also suggest
that the gluon behaves as a massive particle in the low energy
regime and becomes massless in the high energy regime. A
non-vanishing gluon mass is welcome to regularize infrared
divergences and to solve unitarity problems. Furthermore,
a dynamically generated gluon mass is related with the presence of a
$\langle A^2 \rangle$ gluon condensate. A non-vanishing gluon
condensate is a clear sign of the non-perturbative sector of QCD -
see, for example, \cite{Dudal 2004}
and references therein. Besides the theoretical support for a gluon
mass, diffractive phenomena \cite{Forshaw1999} and inclusive
radiative decays of $J/\psi$ and $\Upsilon$ \cite{Field2002} also
suggest a massive gluon.
The precise value for the gluon mass depends on the definition used
to compute it. Lattice QCD simulations and Schwinger-Dyson
equations suggest an infrared gluon mass of $m_g \sim 600$ MeV
\cite{Cornwall1982,Oliveira2011,Oliveira2009}.
Phenomenology points towards a gluon mass between $ 0.5$ GeV and $
1.2$ GeV, depending on how the mass is defined; see table 15 in
\cite{Arriola2004}.

The effective model predicts a constituent light quark mass $M \propto m^2_g$. This result can be
tested looking at the ratio $M / m^2_g$, as a function of momenta, using the decoupling solution
of the SDE. The comparison was performed using the results of \cite{Aguilar2011},
which rely on quenched lattice gluon and ghost propagators to solve
the quark gap equation in the chiral limit. The decoupling solution depends on the ansatz
used for the quark-gluon vertex. In \cite{Aguilar2011} the authors explore two different definitions for
the quark-gluon vertex. It turns out that, in both cases, the ratio of the constituent quark mass over the
effective gluon mass squared is compatible with an essentially constant $M / m^2_g$ in the infrared region.
In this sense, the predictions of our model get some support from the non-perturbative solution of the QCD
Schwinger-Dyson equations.

In the formulation of the effective model we assume flavor independence to reduce the number of independent
parameters. Flavor independence of color interaction has phenomenological implications that have been
checked experimentally at SLAC \cite{Abe1998}, measuring the strong coupling constant $\alpha_s(M_Z)$
in $Z^0 \rightarrow \overline b \, b (g)$, $Z^0 \rightarrow \overline c \, c (g)$ and $Z^0
\rightarrow \overline q_l \, q_l (g)$, where $l = u, d, s$. It was found that, within errors, $\alpha_s(M_Z)$
is flavor independent. Furthermore, flavor independence of strong interactions
was also explored on the theoretical side to understand hadronic phenomena \cite{Beveren1999}.

The effective model adds to the usual QCD  Lagrangian  new operators
which are, from the point of view of the quark operators, either color singlet or
octet.  There is a unique quark singlet operator, which mimics the so-called
$^3P_0$ model \cite{3P0}
used to describe OZI-allowed mesonic strong decays. Recall that the $^3P_0$ model assumes that a pair of
quark--antiquark is created out from the vacuum, with vacuum quantum
numbers, to create new mesons by the recombination of the quarks in
the initial state and the "vacuum quark--antiquark" pair.

%====================================================================
%====================================================================
{\it Effective Degrees of Freedom.}
Quantum Chromodynamics can accommodate multiquark and pure gluon states. From the point of view of an
effective theory, these multiparticle states can be viewed as composite fields.
The success of lattice QCD in the  quenched approximation can be seen as an indication that the non-perturbative
QCD dynamics is mainly in the gluon sector. At the level of the Lagrangian, it is precisely the gluon sector which
contains terms of third and fourth order in the gluon field $A^a_\mu$. We will assume that the non-perturbative
physics is resumed in the simplest tensor structures, i.e. scalar or pseudoscalar fields, whose main
contribution comes from gluonic operators.

For global color transformations $A^a_\mu$ belongs to the octet representation.
Given that $8 \otimes 8 = 1 \oplus 8 \oplus 8 \oplus 10 \oplus \overline{10} \oplus 27$, the
simplest two gluon color operator is a color singlet. Lattice QCD
simulations in the quenched approximation \cite{Cheng2006} predicts
the mass of the low-lying glueballs to be 1710 MeV for quantum
numbers $J^{PC} = 0^{++}$, 2560 MeV for $J^{PC} = 0^{-+}$ and 4780
for $J^{PC} = 0^{+-}$. We aim to build an effective theory describing hadronic phenomena below
the 1 GeV energy scale. From the values for the glueball masses, we expect that the scalar and
pseudoscalar glueballs to play a secondary role and we will not include color singlet scalar fields in
the effective theory.

From the gluon field itself one cannot build a color octet under local transformations.
Instead, one can use the non-abelian Maxwell tensor $F^a_{\mu\nu}$ which,
for local color transformations, belongs to the adjoint representation of $SU(3)$.
The two possible color octet fields with different Lorentz indices are
\begin{equation}
\mathcal{A}_{\mu\nu\eta\zeta} = \frac{1}{\Lambda^3} \,  f_{abc} \, F^b_{\mu\nu}
F^c_{\eta\zeta}
\qquad\mbox{ and }\qquad
\mathcal{S}_{\mu\nu\eta\zeta} =
\frac{1}{\Lambda^3} \, d_{abc}  \, F^b_{\mu\nu}
F^c_{\eta\zeta}  \, , \label{O_def0}
\end{equation}
where $f_{abc}$ ($d_{abc}$) is the antisymmetric (symmetry)
structure constants of $SU(3)$; see, for example, \cite{Leader96} for definitions and
properties. The color transformation properties are derived from $f_{abc}$ and $d_{abc}$
Jacobi identities. Note that in (\ref{O_def0}) one can replace the Maxwell tensor by its dual
without changing its transformations properties.
The composite fields $\mathcal{A}$ and $\mathcal{S}$  were defined
to have the mass dimensions usually associated with a bosonic field.
This requires the introduction of the mass scale $\Lambda$.

The composite operators $\mathcal{A}$ and $\mathcal{S}$
generate multi-gluon states from the QCD vacuum. The contraction of
Lorentz indices for the product of two Maxwell tensors gives rise
either to a scalar field $F^a_{\mu\nu} F^{b \, \mu\nu}$ or a
pseudoscalar field $F^a_{\mu\nu} \widetilde{F}^{b \, \mu\nu}$.
Of course, besides the scalar and pseudoscalar composite fields, other Lorentz tensors can be built from
(\ref{O_def0}). However, we are assuming that the scalars or pseudoscalars
capture the essential of the non-perturbative glue physics and more complex
configurations represented with higher rank tensor operators will be disregarded. The spin zero
composite fields belonging to the color adjoint representation built from equation (\ref{O_def0})  are
\begin{equation}
   \phi^a =  \left\{ \begin{array}{lcl}
   \frac{1}{\Lambda^3}\, d_{abc} \, F^b_{\mu\nu} F^{c \, \mu\nu} \,  , & & \mbox{if } \phi^a \mbox{ is a scalar,} \\
   \frac{1}{\Lambda^3}\, d_{abc} \, F^b_{\mu\nu} \widetilde{F}^{c \, \mu\nu} \,  ,& & \mbox{if } \phi^a \mbox{ is a pseudoscalar  field.}
   \end{array} \right.
\end{equation}
An operator with the same quantum numbers can be built using only quark operators.
Therefore, one can write a scalar color octet composite field as
\begin{equation}
   \phi^a = \frac{a}{\Lambda^3} \, d_{abc} \, F^b_{\mu\nu} F^{c \, \mu\nu} ~ + ~
                 \frac{a^\prime}{\Lambda^2} \big[ \overline q \, t^a q \big] ~ + ~ \cdots \, ,
  \label{general_phi_a}
\end{equation}
where $a$ and $a^\prime$ are constants. The dots in
(\ref{general_phi_a}) stand for other possible operators including
multi-gluon, quarks and ghost fields. If $\phi^a$ is a
pseudoscalar field, in the first term of (\ref{general_phi_a}) one should replace one
of the non-abelian Maxwell tensors by its dual and in the fermion a $\gamma_5$ should
be included.

The definition (\ref{general_phi_a}) allows a na\"{\i}ve estimation of the relative contribution of gluon and quarks to
$\phi^a$ via the ratio $ \mathcal{R} = \langle F^2 \rangle / \Lambda \langle \overline q \, q \rangle$, where $\langle F^2 \rangle$ is the
gluon condensate and $\langle \overline q \, q \rangle$ the light quark condensate. Taking the values
$\alpha_s \langle F^2 \rangle \simeq 0.04$ GeV$^4$
\cite{Narisson2010},
$\langle \overline q \, q \rangle = ( - 270 \mbox{ MeV} )^3$
\cite{Jamin2002,Bordes2010} and
$\Lambda \sim 0.3$ GeV its follows that $\mathcal{R} \simeq 6.8$.
Therefore, one expects $\phi^a$ to be dominated by gluons, in agreement with our initial assumptions.

%===========================================================================
%===========================================================================
{\it The Effective Lagrangian .}
The effective theory includes the gluon field $A^a_\mu$, the quark fields $q_f$, $f$ is a flavor index,
an effective scalar field $\phi^a$ belonging to the adjoint representation of $SU(3)$ color group.
The effective Lagrangian   reads
\begin{eqnarray}
 \mathcal{L} & = & - \frac{1}{4} F^a_{\mu\nu} F^{a \, \mu\nu} ~ + ~
 \sum_f \overline q_f \left\{ i \gamma^\mu D_\mu - m \right\} q_f \nonumber \\
 & &
 + ~ \frac{1}{2} \left( D^\mu \phi^a \right) \left( D_\mu \phi^a \right) - V_{oct}( \phi^a \phi^a ) ~ + ~
  ~ \mathcal{L}_{GF} ~ + ~ \mathcal{L}_{gh} ~ + ~ \mathcal{L}_{new} \,
 \label{new_L}
\end{eqnarray}
where
$D_\mu = \partial_\mu + i g T^a A^a_\mu$  is the covariant derivative,
$T^a$ stands for the generators of $SU(3)$ color group,
$m_f$ the current quark mass associated with flavor $f$,
$V_{oct}$
is the effective potential associated with $\phi^a$.
$\mathcal{L}_{GF}$ is the gauge fixing part of the Lagrangian, $\mathcal{L}_{gh}$ contains the ghost terms and
$\mathcal{L}_{new}$ includes the new interactions. All terms in $\mathcal{L}$ are gauge invariant. The exception
being $\mathcal{L}_{GF}$.
The coupling constant $g$ is not the strong coupling constant.
The effective model assumes that the non-perturbative physics is
summarized in the color octet scalar field $\phi^a$. Therefore, it is expected that the dominant contribution to any
process to be associated with $\phi^a$. The effective gauge coupling constant $g$ parameterizes residual
interactions and it should be a small number, allowing for a perturbative expansion in powers of $g$.

The operators in $\mathcal{L}_{new}$ can be classified accordingly to its mass dimensions.
Given that $\phi^a$ has dimensions of mass, the only gauge invariant dimension four operator that can be
in $\mathcal{L}_{new}$ without breaking parity is
\begin{equation}
    G_4 \sum_f \Big[ \overline q \, t^a \, q \Big] ~ \phi^a \, ,
    \label{O4}
\end{equation}
for a color octet scalar field $\phi^a$. For a pseudoscalar $\phi^a $, $t^a$ should be replaced by $t^a \gamma_5$.
Note that we have considered a single coupling constant $G_4$ to preserve flavor symmetry.
The allowed dimension five operators in $\mathcal{L}_{new}$ are
\begin{eqnarray}
  & &    G_5 \sum_f \Big[ \overline q \, q \Big] ~ \phi^a \phi^a
       + F_1 \sum_f \Big[ \overline q \, t^a  q \Big] ~ d_{abc} \phi^b \phi^c \nonumber \\
  & & +  ~  F_2 \sum_f \Big[ \overline q \, t^a \gamma^\mu q \Big] ~ D_\mu \phi^a +
    F_3 \sum_f \Big[ \overline q \, t^a \gamma^\mu D_\mu q \Big] ~ \phi^a  \, ,
    \label{O5}
\end{eqnarray}
with $G_5 = g_5 / \Lambda$  and
$F_i = f_i / \Lambda$ for $i = 1, 2, 3$ and where $g_5$ and $f_i$ are dimensionless coupling constants.
For a pseudoscalar color octet field only the operators associated with coupling constant $F_1$ and $F_2$ should be
modified by the introduction of a $\gamma_5$ between the quark fields.

The effective field theory is summarized in the Lagrangian   (\ref{new_L}), with $\mathcal{L}_{new}$ given by the
sum of (\ref{O4}) and (\ref{O5}).
$\mathcal{L}$ includes the QCD Lagrangian   and verifies the usual soft-pion theorems of chiral symmetry at low energy.
The new interactions in $\mathcal{L}_{new}$ introduce new vertices not present in the original QCD Lagrangian  ,
namely $q \overline q \phi^a$,
$q \overline q \phi^a \phi^b$ and $q \overline q \phi^a A^b_\mu$, which contribute to quark processes. Note that
the only quark color singlet operator mimics the $^3P_0$ model used to describe OZI-allowed mesonic
strong decays.

So far we have not yet specified the dynamics of the color octet field, i.e.  have not defined its potential energy $V_{oct}$.
However, given that $\phi^a$ cannot give rise to asymptotic $S$-matrix states, for physical processes one has to integrate over $\phi^a$
and it is unnecessary to detailed its dynamics. As discussed bellow, it will be enough to specify its vacuum expectation values.

%=====================================================================================
%=====================================================================================

{\it Dynamical Gluon and Quark masses.}
The kinetic term associated with $\phi^a$ gives rise to a gluon
mass term through the operator
\begin{equation}
  \frac{1}{2}�\, g^2 \, \phi^c (T^a T^b)_{cd} \phi^d A^a_\mu A^{b \, \mu} \, .
\end{equation}
If $\phi^a$ acquire a vacuum expectation value without breaking color symmetry, i.e.
\begin{equation}
  \langle \phi ^a \rangle = 0 \qquad \mbox{ and } \qquad \langle \phi^a \phi^b \rangle = v^2 \delta^{ab} \, ,
\end{equation}
and given that for the adjoint representation $\mbox{tr} \left(  T^a T^b \right)= N_c \, \delta^{ab}$,
then one can write the square of the gluon mass as
\begin{equation}
  m^2_g = N_c \, g^2   v^2 \, ,
  \label{massa_gluao}
\end{equation}
where $N_c = 3$. Note that the condensate $\langle \phi^a \phi^b \rangle$, i.e. $v^2$, and therefore the gluon mass
is gauge invariant. The proof of gauge invariance follows directly from the transformations properties of $\phi^a$.
The gluon mass (\ref{massa_gluao}) is proportional to the effective gauge coupling $g$. As discussed previously, $g$
should be a small number. However, this does not implies necessarily that $m_g$ is also small.
The precise value of $m_g$ depends on the relative values of $v$ and $g$.

The same mechanism shifts the quark masses due to the operator
\begin{equation}
   G_5 \,  \Big[ \overline q \,  q \Big] ~ %\Big(
   \phi^a \phi^a
   \label{anomalous_mass_L}
\end{equation}
giving rise to a constituent quark mass
\begin{equation}
   M_f  ~ = ~ m_f ~ - ~ \left( N^2_c - 1 \right)  \, G_5 \, v^2   %~ - ~ G_5 \, v^2_s
           ~ = ~ m_f ~ - ~ \, \frac{ N^2_c - 1}{N_c} \, \frac{G_5}{g^2} \,  m^2_g           %~ - ~ G_5 \, v^2_s
           \, .
           \label{M_f}
\end{equation}
For light quarks, the constituent mass is given by the quark self energy which, in the model, is linked with the gluon mass,
the second term in equation (\ref{M_f}); note, for our definitions, $G_5$ is a negative number.
If the constituent mass for the light quarks vanishes, chiral symmetry is recovered, and the model predicts that the
gluon mass should also vanish. In this way, the model links chiral symmetry breaking with an effective gluon mass.

\begin{figure}[t] %  figure placement: here, top, bottom, or page
   \centering
   \includegraphics{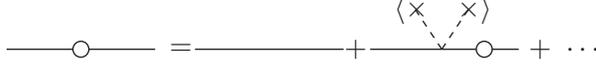}
   \caption{Dressed quark propagator. The dashed lines are associated with the color octet scalar field $\phi^a$.}
   \label{fig:quark_prop}
\end{figure}

%=============================================================
%=============================================================

{\it Light Quark Condensate.}
The ligth quark condensate $\langle \overline q \, q \rangle$ is an order parameter for chiral symmetry breaking,
whose computation requires the knowledge of the quark propagator. In lowest order,
the quark propagator is given by the Feynman graph of figure \ref{fig:quark_prop}, which includes the contribution
from the condensate $\langle \phi^a \phi^b \rangle$. The dressed quark propagator reads
\begin{equation}
   S_0(k) = i \, \frac{\gamma^\mu k_\mu + M}{ ~ k^2 - M^2 ~} \, ,
\end{equation}
where $M$ is the constituent quark mass, and the quark condensate is given by
\begin{eqnarray}
 \langle \overline q \, q \rangle  =
- \, \frac{4 \,  N_c \, N_f}{16 \pi^2} \, M \,   m^2_g ~
 \mathcal{F}\left( \frac{M}{m_g}, \frac{\overline\omega}{m_g} \right),
 \label{light_condensate}
\end{eqnarray}
where $N_f$ is the number of degenerate light flavors, $\overline\omega$ is an energy cut-off and
\begin{equation}
 \mathcal{F}( x, y)  = x^2
       \ln \frac{x^2}{x^2 + y^2} \, + \,  y^2 \, .
\end{equation}
For small $M$ or for $M \ll \overline\omega$, equation (\ref{light_condensate}) becomes
\begin{equation}
 \langle \overline q \, q \rangle ~ = ~ - \, \frac{4 \,  N_c \, N_f}{16 \pi^2} \, M \,   \overline\omega^2 \, .
 \label{light_condensate_M0}
\end{equation}
The quark condensate is proportional to the constituent quark mass, i.e. to the gluon mass squared,
and to the square of the energy cut-off.
A non-vanishing gluon mass means also a non-vanishing quark condensate and, therefore, chiral symmetry is broken.

Equations (\ref{light_condensate_M0}) and $M$ when combined with the results of lattice QCD simulations
\cite{Oliveira2011} allow us to estimate of the various parameters of the theory. Indeed, taking
$\langle \overline q \, q \rangle = \left( - 270 \mbox{ MeV} \right)^3$, $ M = 330$ MeV it follows
from (\ref{light_condensate_M0}) that $\overline\omega = 626$ MeV for $N_f = 2$.
For example, this is a typical value used for the
cut-off when working with Nambu-Jona-Lasinio type of models.
Taking the value for the gluon mass from Landau gauge lattice simulations
$m_g = 648$ MeV, then $G_5/g^2 = -0.295$ GeV$^{-1}$ and
equation (\ref{massa_gluao}) gives $g \, v = 374$ MeV. Note that $g \, v$ is, apart the color factor $\sqrt{N_c}$, the
effective gluon mass and is of order $\Lambda_{QCD}$.

%===========================================================================================
%===========================================================================================

{\it Testing the Gluon and Quark Mass relation.}
The effective model relates the constituent quark mass $M$ and the gluon mass $m_g$ through
equation (\ref{M_f}). For a vanishing current mass, equation (\ref{M_f}) predicts a constant value for
the ratio $M / m^2_g$. This prediction can be tested looking at the solutions of the Schwinger-Dyson equations.
In the following we will use the results published in \cite{Aguilar2011}.
For the gluon and ghost propagators, the authors use the results of lattice QCD simulations and solved the gap equation
for a massless fermion. The calculation does not take into account fermion loops and can be viewed as a quenched approximation.

For the gluon propagator, the inverse lattice propagator was fitted to
\begin{equation}
\Delta^{-1} (q^2) = m^2(q^2) + q^2
   \Big[ 1 + \frac{13 \, C_A \,  g^2_1}{96 \, \pi^2} \ln \left( \frac{q^2 + \rho_1 m^2(q^2)}{\mu^2} \right) \Big] \, ,
\end{equation}
where
\begin{equation}
   m^2 (q^2) = \frac{m^4}{q^2 + \rho_2 m^2}
\end{equation}
and $m = 520$ MeV, $g^2_1 = 5.68$, $\rho_1 = 8.55$ and $\rho_2 = 1.91$ are the fitted parameters,
$\mu = 4.3$ GeV is the renormalization point and $C_A = 3$ is a $SU(3)$ Casimir invariant. The gluon mass is given by
the pole in the propagator and we will use the following definition
\begin{equation}
 m^2_g (q^2) = \frac{ m^2(q^2) }
   { 1 + \frac{13 \, C_A \,  g^2_1}{96 \, \pi^2} \ln \left( \frac{q^2 + \rho_1 m^2(q^2)}{\mu^2} \right) } \, .
\end{equation}

The fermionic gap equation was solved for two different ansatz for the quark-gluon vertex,
namely a non-Abelian improved version of the Ball-Chiu vertex and
an improved version of the Curtis-Pennigton vertex. The choice of vertex leads to slightly different quark mass.
In order to distinguish, the results of the Ball-Chiu vertex will be referred as "BC", while the results from using the
Curtis-Pennington vertex will be referred as "CP".

\begin{figure}[t]
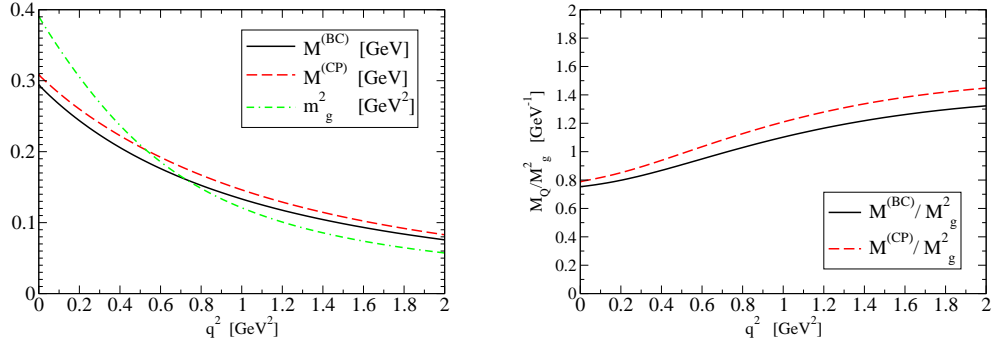
 %  figure placement: here, top, bottom, or page
%\vspace{0.5cm}
   \centering
   \includegraphics[scale=0.25]{masses_q_g.eps}
  \qquad
   \includegraphics[scale=0.25]{mq_over_m2glue_mod.eps}
   \caption{On the left hand side, the plot shows the quark masses from solving the fermionic SDE gap equation, using different
                  ansatz for the quark-gluon vertex, and the squared gluon mass computed from quenched lattice simulations.
                  Note that $M(q^2)$ depends slightly on the definition of the quark-gluon vertex. On the right hand side, the plot
                  shows the ratio $M/m^2_g$.}
   \label{fig:all_masses}
\end{figure}

The quark masses computed using the different vertex ansatz and the squared gluon mass
are reported in figure \ref{fig:all_masses}. For a cut-off of $\overline\omega = 0.626$ GeV, i.e. up to momenta
$q^2 = 0.39$ GeV$^2$, the $M/m^2_g$ ratio increases slightly. Indeed, $M/m^2_g$ changes by less than 15\%,
relative to its zero momentum value, for the BC vertex, and just below 19\% for the CP vertex.

We call the reader attention that the gluon and ghost propagators are quenched propagators and the quark mass was
computed solving the fermionic gap equation using the quenched propagators.
A full dynamical calculation can change the above picture. Lattice QCD simulations \cite{Bowman2005,Silva2010} show a
decreasing constituent quark mass $M(q^2)$ with the current mass in the infrared region.
On the other hand, for dynamical simulations, the gluon propagator is also suppressed relative to the quenched propagator
\cite{Silva2010,Kamleh2007}.
This suggests that $m_g$ is enhanced when we approach the chiral limit.
However, the dynamical ghost propagator is enhanced relative to the quenched propagator \cite{Silva2010} in the infrared
region. An enhancement in the ghost propagator implies an enhancement of the quark-gluon
vertex \cite{Aguilar2011} whose effects remain to be investigated.
Despite the uncertainties related with the simulation of full QCD, hopefully, the scenario observed in figure \ref{fig:all_masses},
where the mass ratio changes by less than 19\%, will remain essentially unchanged. Therefore, one can claim that
the results summarized in figure \ref{fig:all_masses} suggests that $M/m^2_g$ is essentially constant in the
non-perturbative regime. Certainly, given the uncertainties on the calculation of $M$ and $m_g$,
figure \ref{fig:all_masses} will not exclude such a possibility.

%===========================================================================================
%===========================================================================================

{\it Summary.}
We have explored the construction of an effective field theory describing the low energy properties of QCD
using scalar like fields as composite gluon operators.
We found that color singlet scalar-like fields are unlikely to play a dominant role for the infrared properties.
On the other hand, color octet scalar fields allow the building of $SU(3)$ color gauge invariant effective interaction
with interesting properties. The effective model is gauge invariant, flavor symmetric and satisfies the usual current algebra
relations given that it includes the QCD Lagrangian  . It assumes that the non-perturbative physics is resumed
in the new condensate $\langle \phi^a \phi^b \rangle = v^2 \delta^{ab}$, with $v \ne 0$ in the confined phase
and $v = 0$ in the deconfined phase.

Gauge invariance and the scalar condensate are enough to generate a gauge invariant gluon mass, which we associated
with the transverse mass measured in Landau gauge Lattice QCD simulations. $\langle \phi^a \phi^b \rangle$
shifts the quark masses, justifying the difference between current and constituent masses, and generates a light quark condensate
proportional to the constituent quark mass. Further, the model predicts a constant  $M/m^2_g$ ratio.
A massive infrared gluon propagator implies a non-vanishing constituent quark mass and a non-vanishing light quark condensate,
connecting $m_g$  with the breaking of chiral symmetry.
Further, given the relation between $m_g$ and chiral symmetry, the model
predicts that the deconfinement transition and the restoration of chiral symmetry occur simultaneously. This is in good
agreement with the predictions from the studies of the QCD phase-diagram.
If the mass relations are unchanged by temperature and baryonic matter, the model provides a relation between the
critical exponents for the gluon mass $\eta_g$ and the light quark condensate $\eta_q$ near the chiral phase transition, namely
$\eta_q = 2 \, \eta_g$, that can be tested, for example, with lattice QCD simulations.
The mass relation between $M$ and $m_g$ was tested against the decoupling solution of the SDE.

The model includes a number of new interactions. The relations discussed above are the result of exploring two operators,
the kinetic term for the scalar field and the only new color singlet operator at the quark level $\overline q q \, \phi^a\phi^a$. This
last operator mimics the $^3$P$_0$ model used to describe OZI-allowed mesonic strong decays. All the other operators seem
to give marginal contributions to the strong decays because either $\langle \phi^a \rangle = 0$, for the operators multiplying $G_4$, $F_2$
and $F_3$, or $d_{abb} = 0$, for the operator multiplying $F_1$.
The new vertices can contribute the $S$-matrix at the quark level and can give rise to new quark interactions.

In QCD, the quark-gluon vertex is orthogonal to the gluon momenta. The vertices associated with the new interactions do not
verify this property and
can be viewed as mimicking the longitudinal part of the quark-gluon vertex, which is not
constraint by Slavnov-Taylor identities. The effective model has phenomenological implications which have not been explored here
and will be the subject of further publications. For example, the dressing of the quark-photon vertex due to $\phi^a$ can give
rise to quark anomalous magnetic moments of the type discussed in \cite{Chang2011}.

The model building blocks are the color octet field $\phi^a$, the quark and the gluon fields. The
hadronic matter is essentially quark matter. The experimental situation for pure glue states is not so clear.
One has to understand why there are no physical states than can be associated with $\phi^a$. An explanation could
be its low mass and the weak couplings to the quarks.
 Looking at the Lagrangian, one can estimate the contribution of the new terms to the mass of
$\phi^a$ as $\delta m^2_\phi \approx | G_5 \, \langle \overline q \, q \rangle |$. For
$G_5 = - 0.295 \, g^2$ GeV$^{-1}$ and  $\langle \overline q \, q \rangle = ( -270 \mbox{ MeV})^3$, it follows that
$\delta m_\phi \lesssim 76$ MeV. Curiously, recently the authors of \cite{Beveren2011} have analyzed the BABAR
bottomonium decay data and found an excess of signal which could be explained by a light scalar particle with a mass
of 38 MeV.

Finally, we note that the model, being a valid description for low energy QCD, predicts a behavior for the infrared gluon propagator
which distinguish between the so-called scaling solutions and decoupling solutions of SDE. The model is in close agreement with
the decoupling like solution.

%==========================================================
%==========================================================
\section*{Acknowledgements}

The authors acknowledge financial support from the Brazilian
agencies FAPESP (Funda\c c\~ao de Amparo \`a Pesquisa do Estado de
S\~ao Paulo) and CNPq (Conselho Nacional de Desenvolvimento
Cient\'ifico e Tecnol\'ogico). OO acknowledges financial support from FCT under
contract PTDC/FIS/100968/2008.

%==========================================================
%==========================================================

\end{document}